\documentclass[9pt,twocolumn,twoside]{optica}
\setboolean{shortarticle}{false}
\setboolean{minireview}{false}
\usepackage{microtype} 						
\usepackage{amsmath} 							
\usepackage{amssymb}							

\renewcommand{\b}[1]{\mathbf{ #1}}				
\newcommand{\h}[1]{\hat{ #1}}					
\renewcommand{\d}{\mathrm{d}}			       
\newcommand{\ket}[1]{| #1 \rangle}				
\newcommand{\m}[1]{\langle  #1 \rangle}				
\newcommand{\Tr}{\mathrm{Tr}}			       

\newcommand{\odd}{\mathrm{odd}}				
\newcommand{\even}{\mathrm{even}}				

\title{Quantum optical feedback control for creating strong correlations in many-body systems}

\author[1,*]{Gabriel Mazzucchi}
\author[1,2]{Santiago F. Caballero-Benitez}
\author[3]{Denis A. Ivanov}
\author[1]{Igor B. Mekhov}

\affil[1]{Department of Physics, Clarendon Laboratory, University of Oxford, Parks Road, Oxford OX1 3PU, United Kingdom}
\affil[2]{Instituto de F\'{\i}sica, Universidad Nacional Aut\'onoma de M\'exico, Apartado Postal 20-364, 01000 Ciudad de M\'exico, M\'exico}
\affil[3]{St. Petersburg State University, Ulianovskaya 3, Petrodvorets, St. Petersburg, 198504, Russia}

\affil[*]{Corresponding author: gabriel.mazzucchi@physics.ox.ac.uk}

\dates{Compiled \today}

\ociscodes{(020.1335) Atom optics; (270.0270) Quantum optics; (020.1475) Bose-Einstein condensates; (020.1670) Coherent optical effects; (270.5580) Quantum electrodynamics; (270.5570) Quantum detectors.}

\doi{\url{http://dx.doi.org/10.1364/optica.XX.XXXXXX}}

\begin{abstract}
Light enables manipulating many-body states of matter, and atoms trapped in optical lattices is a prominent example. However, quantum properties of light are completely neglected in all quantum gas experiments. Extending methods of quantum optics to many-body physics will enable phenomena unobtainable in classical optical setups. We show how using the quantum optical feedback creates strong correlations in bosonic and fermionic systems. It balances two competing processes, originating from different fields: quantum backaction of weak optical measurement and many-body dynamics, resulting in stabilized density waves, antiferromagnetic and NOON states. Our approach is extendable to other systems promising for quantum technologies.
\end{abstract}

\setboolean{displaycopyright}{true}

\begin{document}

\maketitle
\thispagestyle{fancy}
\ifthenelse{\boolean{shortarticle}}{\abscontent}{}

\section{Introduction}

Optics provides a very high degree of control and measurements in various systems. This has enabled observations of such fundamental building blocks of quantum physics as entanglement and measurement-induces collapse of the wave function (the so-called quantum measurement backaction). These phenomena have been already demonstrated in various spectral ranges of the elctromagnetic radiation, where the backaction was an essential ingredient for preparing Fock and Schr\"odinger cat states \cite{HarocheBook,Vlastakis2013, Reiserer2013,McConnell2015}. However, the observation of such delicate quantum effects in genuinely many-body systems with strong correlations remains a challenge. One of the most prominent examples of well-controllable many-body systems is ultracold atoms trapped in periodical micropotentials created by laser beams: optical lattices. Quantum gases are a unique tool that can be used for studying important  effects originating from different disciplines, including quantum information, condensed matter, and high energy physics~\cite{Lewenstein}. Although the light plays a crucial role in these setups, its quantum nature is largely neglected in the experiments so far: the light is used as a classical tool for creating intriguing many-body states of atoms. In contrast, elevating light to a fully quantum variable will enrich many-body physics and allow novel phenomena and techniques at the ultimate quantum regime of the light-matter interaction (see \cite{Mekhov2012,ritsch2013} for reviews). In particular, the quantum backaction of weak (non-fully projective) optical measurement can be introduced in the many-body systems, where it constitutes a novel source of competitions \cite{Mazzucchi2016}.  

Here we show how the use of quantum optical techniques can push forward the engineering of many-body strongly correlated states, which are hardly obtainable in setups using classical light beams. Specifically, we focus on the feedback control of ultracold bosonic and fermionic atoms in optical lattices, subjected to the quantum weak measurement. Very recent experiments~\cite{Hemmerich2015,Esslinger2015} succeeded in coupling atoms in lattices to high-Q optical cavities, realizing for the first time realistic systems for our proposal. This have already led to the detection of new quantum phases where the interaction between the atoms is mediated by the light field~\cite{Moore1999, Chen2009, Caballero2015, Caballero2015a,Caballero2016,Niedenzu2013,Caballero2016b}. While the dynamical nature of light is very important in these cavity QED experiments, its quantum properties still require exploitation, which is a focus of our paper. Using quantized light will open the possibility of realizing several proposals for quantum nondemolition (QND) measurements~\cite{MekhovPRA2009, Roscilde2009, Rogers2014, Eckert2007, HaukePRA2013, Rybarczyk2015, Atoms,Ashida2015}, which rely on the entanglement between matter and light field for mapping the correlations between the atoms on the scattered light and detecting eventual phase transitions \cite{mekhovLP2009,mekhovLP2010,mekhovLP2011,Kozlowski}.

In this paper we will show that  combining the effects arising from measurement backaction and feedback, it is possible to realize in real-time~\cite{Haroche2011} stable strongly correlated many-body states such as density waves, NOON, and antiferromagnetic states (even in the absence of atomic interactions), which are important for simulating analogous condensed matter systems and have applications in quantum information. In contrast to recent proposals~\cite{Pedersen2014,Ruostekoski2014,Ivanov2014,Ivanov2016}, where measurements are performed at optimized moments in time or the specific series of quantum jumps is approximated with a continuous signal, we control the quantum state of the system by modulating the system parameters in a single quantum trajectory depending on the outcome of quantum weak measurement. Quantum optical control allows us not only create, but also tune the stationary and dynamical parameters of the many-body states (e.g. the particle imbalance and oscillation frequency).   
 
\begin{figure}[t!]
\includegraphics[width=.45\textwidth]{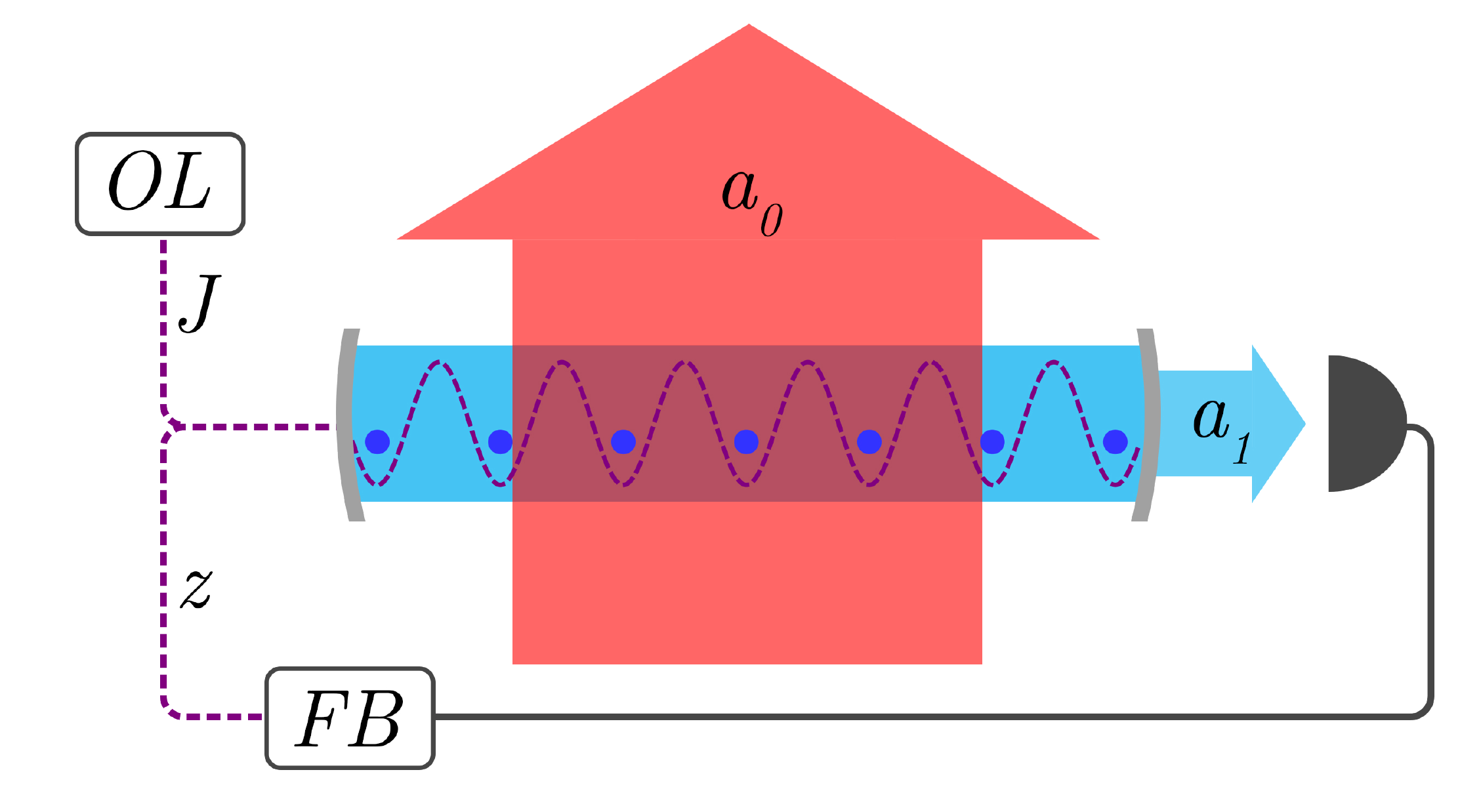}
\caption{Experimental setup. Ultracold atoms are loaded in an opical lattice (OL) inside an optical cavity and probed with a coherent light beam (mode $a_0$). The cavity enhances the light scattered orthogonally to $a_0$ and the photons escaping it are detected (mode $a_1$). Depending on the measurment outcome, a feedback (FB) loop with gain $z$ is applied for modulating the depth of the optical lattice.} \label{fig:setup}
\end{figure}
\begin{figure*}[t!]
\includegraphics[width=.95\textwidth]{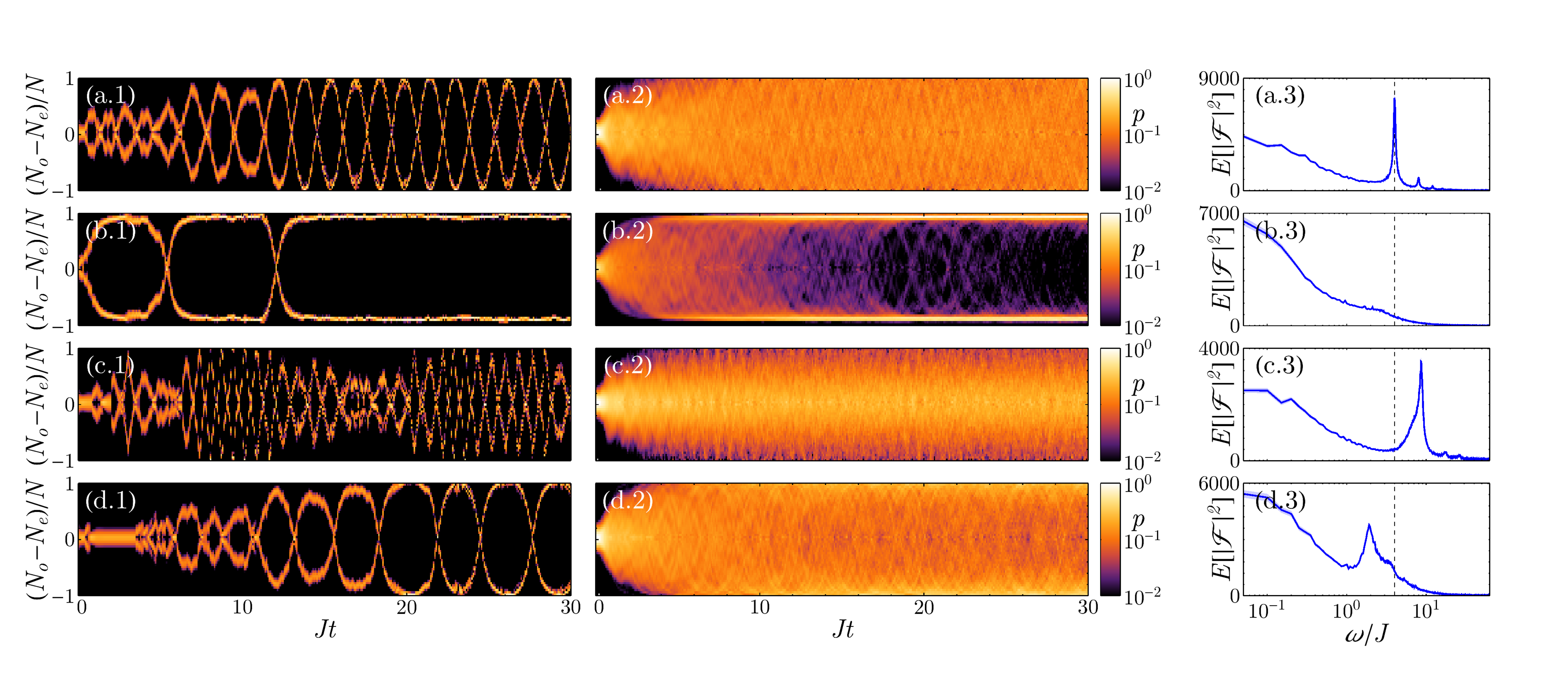}
\caption{Probability distribution of $\h{N}_\odd-\h{N}_\even$ for single quantum trajectories (Panels 1) and averages over 200 trajectories (Panels 2) for different values of the feedback gain $z$. Panel 3 shows the power spectrum of  $\m{\h{N}_\odd-\h{N}_\even}$ averaged over 200 trajectories. In the absence of feedback [Panel (a), $z=0$] the oscillations of the population of the odd sites are visible only in a single trajectory. For $z>z_c$ [Panel (b), $z=1.23 z_c$] the imbalance between odd and even sites is frozen for each quantum trajectory and $E \left[ |\mathcal{F}|^2\right]$ does not have a strong peak, indicating that $\h{N}_\odd-\h{N}_\even$ does not oscillate. For $z<z_c$ the  frequency of the oscillations can be tuned above [Panel (c)  $z=-4 z_c$] or below  [Panel (d) $z=0.8 z_c$] the frequency defined by the tunnelling amplitude $J$. Again, the oscillatory dynamics is visible only in a single quantum trajectory and the average probability distribution spreads quickly. ($N=100$, $\gamma/J=0.02$, $J_{jj}=(-1)^j$, $z_c=0.0025$)}\label{fig:fig2}
\end{figure*}

The effects presented in this work can be generalized to other many-body (or simply multimode) physical systems such as optomechanical arrays \cite{Paternostro2011,Aspelmeyer2014}, superconducting qubits as used in circuit cavity QED \cite{Palacios2010,Paraoanu2011,Pirkkalainen2013,White2015}, and even purely photonic systems (i.e. photonic chips or circuits) with multiple path interference, where, similarly to optical lattices, the quantum walks and boson sampling have been already discussed \cite{Spring2013,Nitsche2016,Brecht2015,Elster2015,Oren:16}. Recently it has been achieved ultra-strong light matter coupling in a 2D electron gas in THz metamaterials~\cite{Zhang2016}, while developments have been made with respect to light induced  high-Tc superconductivity in real materials \cite{Cav1,Mitrano,Dieter1}. This further opens the possibility to engineer what we propose in real solid state materials and hybrid devices for quantum technologies \cite{MolmerHyb} in the near future.

\section{Theoretical model}

We consider a system of ultracold atoms loaded in an optical lattice inside an optical cavity with decay rate $\kappa$ (see Figure~\ref{fig:setup}). A classical light beam illuminates the atomic ensemble and scatters photons inside the cavity, which allows to enhance the light in a particular direction~\cite{Bux2013,Kessler2014,Landig2015}. The atomic dynamics is described by the tunneling Hamiltonian $\h{H}_0=- \hbar J \sum_{\langle i, j \rangle} b^{\dagger}_i b_j$ where $J$ is the tunneling amplitude between nearest neighbors sites of the optical lattice and the operator $b^{\dagger}_i$ ($b_i$) creates (annihilates) an atom at the site $i$.  Because of the interaction, light and matter become entangled. Similarly to the well-known Einstein-Podolsky-Rosen (EPR) effect, detecting one of  the entangled subsystems (photons escaping the cavity) affects another one (atoms), manifesting the quantum measurement backaction. This allows to probe and modify the quantum state of atoms via the quantum measurement. Furthermore, the information from the photodetections is used for applying feedback to the system, dynamically stabilizing interesting quantum states that can be targeted and obtained deterministically. As in classical optics, the amplitude of the far off-resonant scattered light is proportional to the atomic density and it follows $a_1 \sim \int \! u_{\mathrm{out}}^*(\b{r})  u_{\mathrm{in}}(\b{r}) \h{n}(\b{r})  \d \b{r}$ where $ n(\b{r})$ is the atomic density operator, and $u(\b{r})$ are the mode functions of the probe and scattered light modes. Introducing the Wannier functions $w(\b{r})$ of the optical lattice and neglecting the contribution from the inter-site density, the operator $a_1$ reduces to $a_1=C \h{D}$ where 
\begin{align}\label{defC}
C=\frac{i \Omega_{10} a_0}{i \Delta_p - \kappa},
\end{align}
is the Rayleigh scattering coefficient~\cite{Elliott2015,Mazzucchi2016,Atoms}, $a_0$ is the amplitude of the (classical) coherent probe, $\Omega_{10}=g_1 g_0/\Delta_a$, $g_l$ are the atom-light coupling constants, $\Delta_a$  and  $\Delta_p$ are respectively the atom-light and probe-cavity detunings, $\h{D}=\sum_j J_{jj} \h{n}_j$, $\h{n}_j=b^\dagger_j b_j$ is the density operator for the lattice site $j$ and $J_{jj}=\int w^2(\b{r}-\b{r}_j) u_{\mathrm{out}}^*(\b{r})  u_{\mathrm{in}}(\b{r})   \d \b{r}$. For fermions, different (circular) light polarizations couple to different spin states~\cite{Meineke2012, Sanner2012}: we exploit this property for probing linear combinations of the spin-$\uparrow$ and spin-$\downarrow$ atomic density changing the polarization of the probe beam. Focusing on the case of linearly polarized light ($a_{1x}$ and $a_{1y}$), we find that the measurement is sensitive to the local density  $a_{1x}=C \h{D}_x=C \sum_j J_{jj} \h{\rho}_j$ ($\h{\rho}_j=\h{n}_{j \uparrow} + \h{n}_{j \downarrow}$) or the local magnetization $a_{1y}=C \h{D}_y=C \sum_j J_{jj} \h{m}_j$ ($\h{m}_j=\h{n}_\uparrow - \h{n}_\downarrow$). 

Crucially, the couplings $J_{jj}$ can be engineered by changing the mode functions of the light modes and/or the angle between the probe beam, the scattered beam and the optical lattice. This allows us to probe different global quantities characterizing the quantum state of the atoms which can be used for quantum state engineering. Moreover, if $J_{jj}$ has the same value on a several sites of the optical lattice, the light scattering partitions the optical lattice in macroscopically occupied spatial modes where the atoms scatter light with the same intensity and phase, making them indistinguishable to the measurement. In this work, we focus on two detection schemes. If the light modes are standing waves and their interference pattern has nodes at the odd sites of the lattice, one has $J_{jj}=0$ for odd and $J_{jj}=1$ for even sites. Therefore, the measurement probes the number of atoms occupying the even sites of the lattice and induces the formation of two spatial modes defined by lattice sites with different parity. The same mode structure can be obtained considering traveling or standing waves such that the wave vector of the cavity ($\b{k}_\mathrm{out}$) is along the lattice direction and the wavevector of the probe ($\b{k}_\mathrm{in}$) is orthogonal to it so that $(\b{k}_\mathrm{in}-\b{k}_\mathrm{out}) \cdot \b{r}_j=\pi j$. This configuration corresponds to detecting the photons that are scattered in the diffraction minimum at 90$^\circ$ and $J_{jj}=(-1)^j$, i. e. the jump operator is sensitive to the population difference between odd and even sites.

We focus on the outcome of a single experimental run and describe the conditional evolution of the atomic state using the quantum trajectories formalism. In general, the dynamics of a system subjected to continuous monitoring and feedback follows the master equation \cite{Wiseman}
\begin{equation}\label{eq:master}
\d \h{\rho} (t)= \left\{   \d N \left[ \mathrm{e}^{\mathcal{K}} \left (\mathcal{G} [ \hat{c} ]+1 \right) -1 \right] - \d t \mathcal{H} [i \hat{H}_0 + \frac{1}{2} \hat{c}^\dagger \hat{c} ] \right\}\h{\rho}(t)
\end{equation} 
 where $\h{c}=\sqrt{2 \kappa} a_1$ is the jump operator, $\d N$ is the stochastic It\^o increment such that $E[\d N]= \Tr[\hat{c} \h{\rho} \hat{c}^\dagger]\d t$,  $\mathcal{G}$ and  $\mathcal{H}$ are the superoperators
\begin{align}
\mathcal{G} [\hat{A} ]\h{\rho}=\frac{\hat{A}\rho\hat{A}^\dagger }{\Tr \left[\hat{A}\h{\rho}\hat{A}^\dagger \right]}-\h{\rho}\\
\mathcal{H} [\hat{A} ]\h{\rho}=\hat{A} \h{\rho}+ \h{\rho} \hat{A}^\dagger - \Tr\left[ \hat{A} \h{\rho} +\h{\rho} \hat{A}^\dagger \right],
\end{align}
$\h{H}_0$ describes the coherent (free) evolution of the system and the feedback loop acts on the master equation with delay $\tau$ as $
\left[ \h{\rho}(t+\d t) \right]_{\mathrm{fb}}=  \exp \left[  \d N(t - \tau) \mathcal{K}  \right] \h{\rho}(t)$.

We consider the case where the feedback loop changes the depth of the atomic potential instantaneously ($\tau \rightarrow 0$), effectively modulating the value of the tunneling amplitude $J$ depending on the photocount rate. Within these assumptions, the superoperator $\mathcal{K}$ acts on the density matrix as 
\begin{align}
\mathcal{K} \h{\rho} = i [z\h{H}_0,\h{\rho}]
\end{align}
where the parameter $z$ describes the feedback strength. Assuming perfect detection efficiency and that the initial state of the system is in a pure state, we solve the master equation \eqref{eq:master} by simulating individual quantum trajectories. The evolution of the system is determined by the stochastic process described by the quantum jump operator $\h{d}=\sqrt{2 \kappa} \mathrm{e}^{i z \h{H}_0}\h{c}$ and the  non-Hermitian Hamiltonian $\h{H}_\mathrm{eff}= \h{H}_0 - i \hbar  \hat{c}^\dagger  \hat{c}/2$. In other words, $\h{H}_\mathrm{eff}$ generates the dynamics of the atomic system in between two consecutive photoemissions and,  when a photon escapes the optical cavity, the jump operator $\h{d}$ is applied to the atomic wavefunction. Note that $\h{d}$ describes both the effects of measurement backaction ($\h{c}$) and the feedback loop ($\mathrm{e}^{- i z \h{H}_0}$). Finally, we characterize the strength of the measurement process with the ratio $\gamma / J$ where $\gamma=\kappa |C|^2$: this quantity determines if the dynamics of the system in the absence of feedback is dominated by the tunneling or by the quantum jumps. 

\section{Feedback-stabilized density waves and antiferromagnetic ordering}

We first consider non-interacting bosons and demonstrate that the feedback process can be used for targeting specific quantum states even in the weak measurement regime ($\gamma \ll J$). Collecting the photons scattered in the diffraction minimum, the detection scheme probes the population imbalance between the odd  and even sites of the lattice ($\h{c}\propto \h{N}_{\odd}- \h{N}_{\even}$).  Importantly, since the intensity of the scattered light is $a^\dagger a$, the measurement is not sensitive to the sign of the imbalance and the atomic state remains in a superposition of states with opposite imbalance (Schr\"odinger cat state). Moreover, in the absence of feedback, the measurement backaction induces giant oscillations in the atomic population \cite{Mazzucchi2016,Mazzucchi2016b} which resemble a dynamical supersolid state \cite{Caballero2015} (Figure \ref{fig:fig2}a). Interestingly, these oscillations are visible only in a single experimental run (and not in average quantities), since their phase varies randomly between different quantum trajectories. Nevertheless, the oscillations are fully visible in the measured signal. We determine the frequency of such oscillations by computing the power spectrum of the photocurrent $\m{a^\dagger a} (t)$ for a single quantum trajectory: this quantity is directly accessible in the experiments and, being proportional  to $\m{(\h{N}_{\odd}- \h{N}_{\even})^2}$, allows to estimate the absolute value of the difference in population between odd and even sites. In order to characterize the behavior of all the trajectories, we calculate $\mathcal{F}=\int \m{(\h{N}_{\odd}- \h{N}_{\even})^2} e^{- i \omega t} \d t$  for each realization of the conditional evolution and we then average  $|\mathcal{F}|^2$ over several quantum trajectories ($E \left[ |\mathcal{F}|^2\right]$).  As expected, if the feedback is not present ($z=0$), $E \left[ |\mathcal{F}|^2\right]$ has a strong peak at $\omega= 4 J $ indicating that the oscillation frequency is determined by the tunneling amplitude.

\begin{figure}[t!]
\includegraphics[width=.5\textwidth]{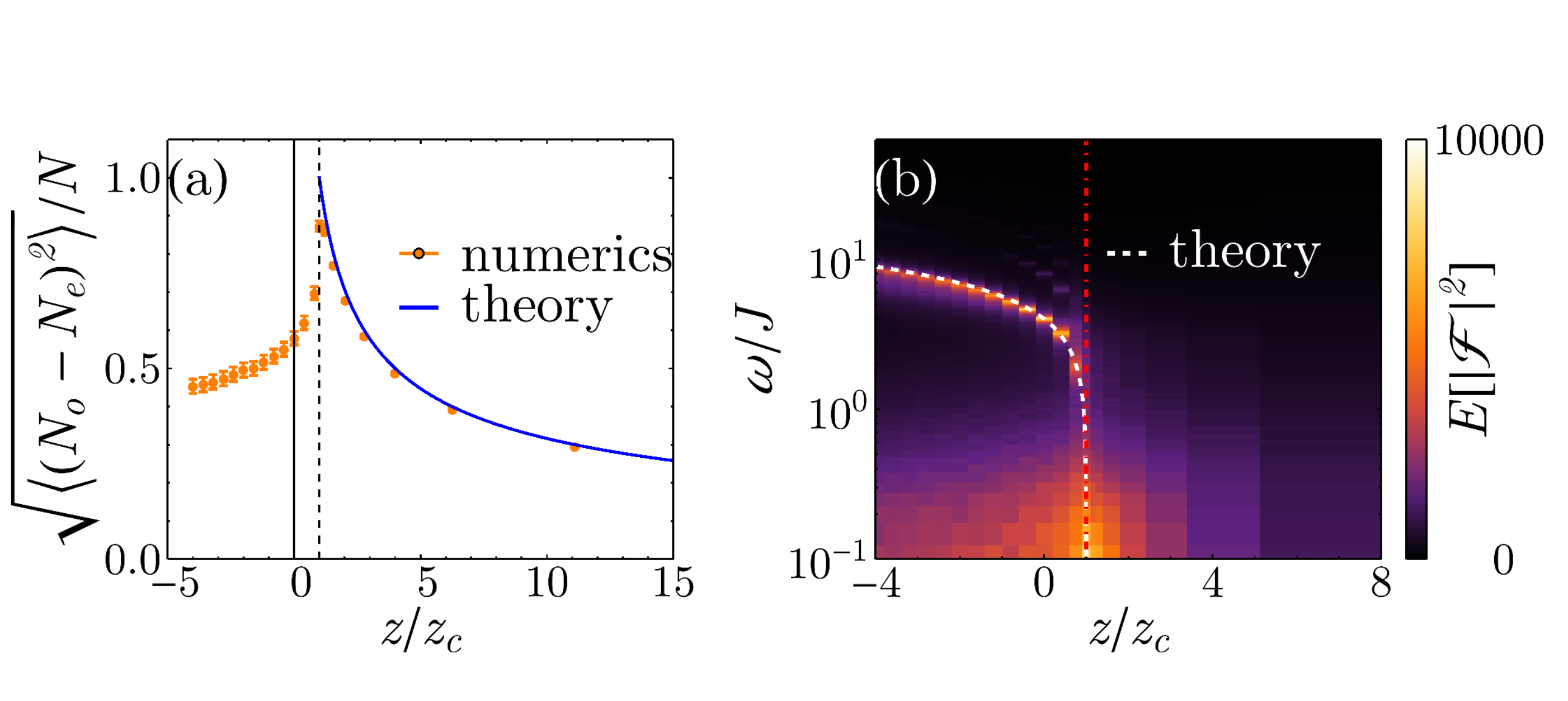}
\caption{Effects of measurement and feedback detecting the photons scattered in the diffraction minimum. Panel (a): Imbalance between odd and even sites as a function of the feedback strength. There is a very good agreement between the numerical results and the analytic expression derived in text.  Panel (b): average power spectrum as a function of the feedback strength. The value of  $E \left[ |\mathcal{F}|^2\right]$ presents a strong peak for $z<z_c$, indicating that the trajectories are characterized by an oscillatory dynamics. The vertical dashed line marks $z=z_c$. ($N=100$, $\gamma/J=0.02$, $J_{jj}=(-1)^j$, $z_c=0.0025$) } \label{fig:cfr1}
\end{figure}

Considering now the case when feedback is applied to the system, we find that there is a critical value for the parameter $z_c$ which defines two different dynamical regimes: if $z<z_c$ the expectation value $ \m{(\h{N}_{\odd}- \h{N}_{\even})^2}$ oscillates  while if $z>z_c$ the imbalance reaches a steady state value that is deterministically defined by the parameter $z$ itself (Figure \ref{fig:fig2}). We explain this effect by looking at the effect of feedback on atoms: defining $\Delta t_n$ to be the time interval between the $(n-1)-$th and $n-$th quantum jump, the state of the system after $N_{ph}$ photocounts is
\begin{equation}
\ket{\psi(t;N_{ph})}\propto\prod_{n=2}^{N_{ph}} \left[\mathrm{e}^{- i \h{H}_\mathrm{eff} \Delta t_n}\mathrm{e}^{i z \h{H}_0} \h{c} \right] \mathrm{e}^{- i \h{H}_\mathrm{eff} \Delta t_1} \ket{\psi_0}
\end{equation} 
where $\ket{\psi_0}$ is the initial state of the system. In the weak measurement regime ($\gamma \ll J$), we can focus on the terms depending linearly on $ \Delta t_n$ or $z$ and  neglect the commutators between $\h{H}_\mathrm{eff}$ and $\h{H}_0$ since it scales as $ \Delta t_n z$, thus
\begin{equation}
\mathrm{e}^{- i \h{H}_\mathrm{eff} \Delta t_n}\mathrm{e}^{i z \h{H}_0} \approx \mathrm{e}^{-i( \Delta t_n-z)\h{H}_0- \hbar \h{c}^\dagger \h{c} \Delta t_n /2}.
\end{equation} 
Therefore, the parameter $z$ defines an effective timescale which competes with the tunneling and the measurement processes. We find that it is possible to formulate a simple description of the dynamics of the atomic system by comparing the value of $z$ to the average time interval between two consecutive quantum jumps, i. e. $\overline{\Delta t}=1/(2 \gamma \m{\h{D}^\dagger \h{D}})$. Specifically, if $\Delta t_n \approx\overline{\Delta t}= z$, the feedback completely inhibits the tunneling  described by the Hamiltonian $\h{H}_0$ and the dynamics of the system is determined by the ``decay'' term in $\h{H}_\mathrm{eff}$. In this case, there are only two processes which contribute to the evolution of the system: the non-Hermitian dynamics (which tends to suppress the atom imbalance) and the quantum jumps (which drive the system towards large $\m{(\h{N}_{\odd}- \h{N}_{\even})^2}$). In the large time limit, these two effects balance each other and the system reaches a steady state where the probability distribution of $\h{N}_{\odd}- \h{N}_{\even}$ has two narrow peaks at opposite values. This regime is somehow analogous to the strong measurement regime (quantum Zeno effect) \cite{Kozlowski2015NH,KozlowskiBond} and to the one described in previous works where the atomic dynamics was neglected. However, in these cases the system is confined to a quantum state that is an eigenvector of the jump operator whose eigenvalue is determined stochastically and ultimately depends on the initial state of the system. In contrast, here the introduction of a feedback loop allows us to deterministically select the final state of the system by tuning the value of $z$. Defining $\m{\h{D}^\dagger \h{D}}_{T}$ as the target imbalance one wants to obtain, the corresponding feedback gain realizing this specific configuration is $z=1/(2 \gamma \m{\h{D}^\dagger \h{D}}_{T})$ [see Figure~\ref{fig:cfr1}(a)]. Moreover, the target steady state is reached independently from the initial state of the system.

Since the value of the population imbalance between odd and even sites cannot exceed the total number of atoms, the maximum possible value for $\m{\h{D}^\dagger \h{D}}_{T}$ is $N^2$. This defines a critical $z$ under which the condition $\overline{\Delta t}= z$ cannot be fulfilled, $z_c=1/(2 \gamma N^2)$. As a consequence, for $z<z_c$ the state of the system does not reach a steady state and the measurement backaction establishes an oscillatory dynamics. Following the approach presented in \cite{Mazzucchi2016b}, we find that carefully choosing the feedback gain it is possible to tune the frequency of the oscillations of $\m{(\h{N}_{\odd}- \h{N}_{\even})^2}$  according to $\omega=4 \sqrt{1-z/z_c}$ [see Figure~\ref{fig:cfr1}(b)]. Again, these oscillations are visible only analyzing single quantum trajectory and are not visible in the average probability distribution. The presence of two peaks in the probability distribution  $\m{\h{N}_{\odd}- \h{N}_{\even}}$ makes this measurment setup susceptible to decoherence due to photon losses~\cite{MekhovPRA2009}. However, this scheme can be made more robust by illuminating only the odd sites of the lattice so that $\h{c}\propto\h{N}_{\odd}$. In this case, the measurement operator probes the occupation of the odd sites and its probability distribution has only one strong peak.

\begin{figure}[t!]
\includegraphics[width=.45\textwidth]{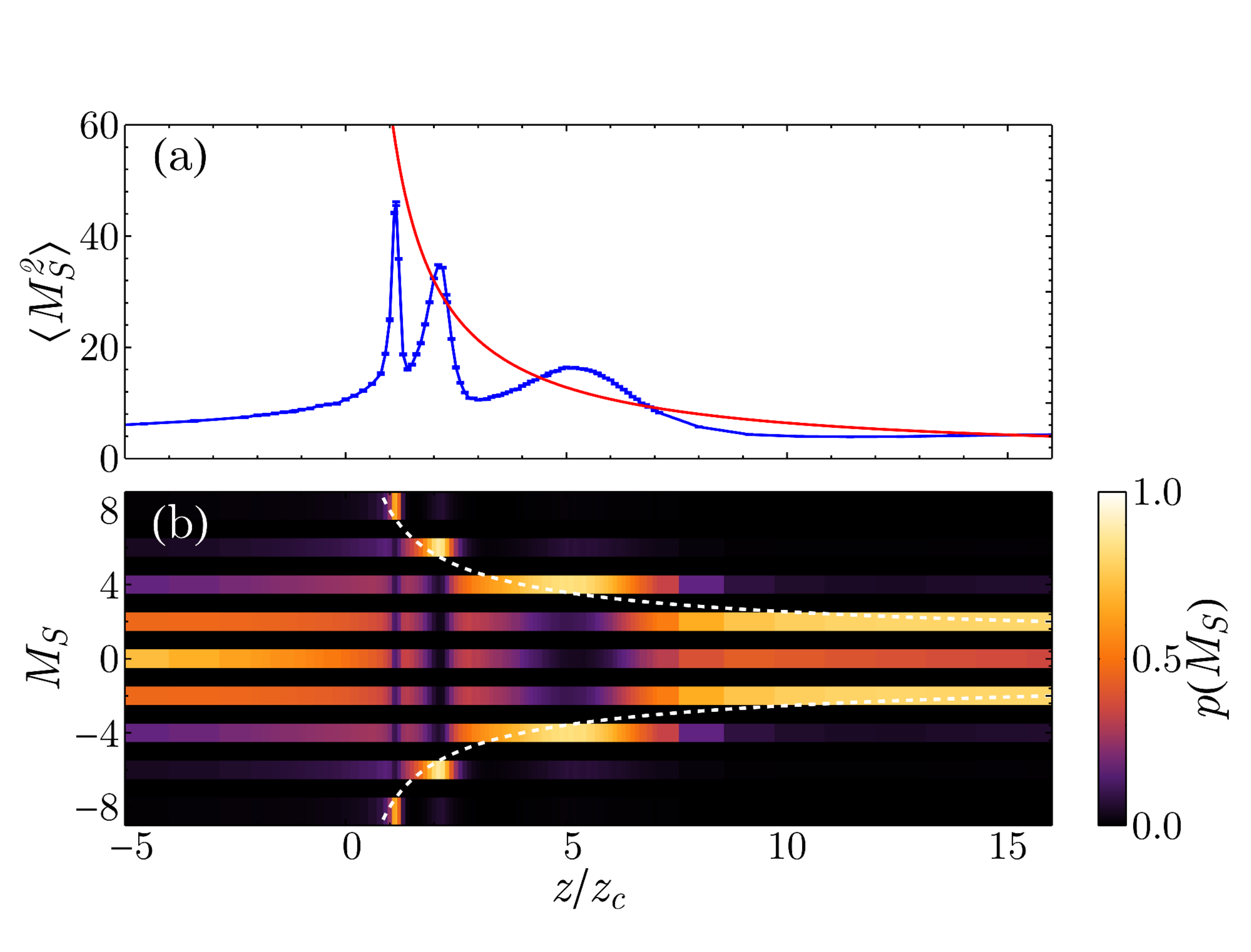}
\caption{Effects of measurement and feedback probing the staggered magnetization in a fermionic system. (a): square of the staggered magnetization as a function of the feedback strength (blue line) compared to the analytic formula (red line) for a fermionic system. Note that the two curves do not have the same behavior because the analytic solution assumes that $\h{M}_S$ is a continuous variable while the numerical simulations are performed on a small system where $\h{M}_S$ assumes only discrete values. Panel (b): steady state value of the probability distribution of $\h{M}_S$ as a function of the feedback strength. The dashed line represents the theoretical prediction. The feedback loop stabilizes antiferromagnetic correlations. ($N_\uparrow=N_\downarrow=4$, $\gamma/J=1$, $J_{jj}=(-1)^j$ , $z_c=1/128$)} \label{fig:fermi}
\end{figure}

We now turn to non-interacting fermions and we focus on the case where linearly polarized photons are detected so that the jump operator is sensitive to the staggered magnetization $\h{M}_S=\h{M}_\odd - \h{M}_\even$. If feedback is not present, the measurment backaction leads to quantum states characterized by antiferromagnetic ordering \cite{Mazzucchi2015}. However, these correlations follow an oscillatory dynamics and cannot be selected deterministically since they are a result of the competition between local tunneling processes and the (stochastic) quantum jumps. In analogy to the bosonic case, introducing a feedback loop allows us to obtain antiferromagnetic states with a predetermined staggered magnetization in each single quantum trajectory even in the absence of many-body interactions. Again, there is a critical value of the gain $z$ which sharply divides two regimes. If $z>z_c$ the system reaches a steady state such that $\m{\h{M}_S}=\sqrt{1/(2 \gamma z)}$ for each quantum trajectory. In contrast, if $z<z_c$ the value of  $\m{\h{M}}_S$ is not stationary and taking its expectation over many quantum trajectories we find that, on average, the atomic state does not present antiferromagnetic order. Figure~\ref{fig:fermi} illustrates this effect by showing the average over many realizations of the expectation value of the staggered magnetization and its probability distribution in the large time limit  as a function of the feedback gain. Note that the predicted value for $\h{M}_S$ agrees with the numerical results only qualitatively. This is because the analytic solution $\m{\h{M}_S}=\sqrt{1/(2 \gamma z)}$ treats the staggered magnetization as a continuous variable while  when performing a simulation on a small system only some discrete values of $\m{\h{M}_S}$ are possible. This effect is not surprising and it is rather analogous to previous works where the discreteness of the matter field leads to spectra with multiple peaks \cite{BruneDiscrete,GambettaDiscrete,MekhovDiscrete}. 

\section{Generalization to other physical systems}

In this work we focused on the effect of feedback on the conditional dynamics of an ultracold atomic system. However, the effects we described can be generalized to other physical systems where the measurement backaction can compete with the usual unitary dynamics. For example, the phenomena we presented could be observed in purely photonic systems with multiple path interferometers known as photonic chips or photonic circuits \cite{Spring2013, Holleczek2015,Oren:16,Marchildon:16}. In these systems, single photons can propagate across multiple paths, generating dynamics which is analogous to the tunneling of atoms in an optical lattices. Moreover,  the propagation of the photons that can be easily controlled by tuning the reflection and transmission coefficients of the beam splitters or waveguide couplers, allows to implement feedback \cite{Nitsche2016} on the system. Such setups are within the reach of current experiments where single photons and photonic pairs have been successfully used as input states for multiple waveguides \cite{Spring2013}. The effects demonstrated in such photonic systems include quantum walks  \cite{Eichelkraut2014,Elster2015} and boson sampling  \cite{Spring2013}.

The non-destructive detection of photons is very difficult to implement and it is certainly a challenge for current experiments. However, there are some techniques which allow to perform this task. For example, measuring an idler beam arising from the creation of pairs of entangled photons or beams via parametric down conversion represents a QND measurement of the signal beam~\cite{Wiseman}. Moreover, such setup is consistent with current experiments involving photon pairs~\cite{Spring2013}. Another possibility for realizing QND of photons is to use a cavity QED system~\cite{Reiserer2013}. If the number of photons participating to the multiple paths interference is not too small, the usual destructive detection of few of them can be considered as a measurement that is not fully projective since the surviving photons continue to propagate. This is analogous to the creation of nonclassical states of light and quantum correlations using photon subtraction~\cite{Paternostro2011}. 

\section{Conclusions}

In conclusion, we have shown that using techniques from quantum optics can broaden the field of many-body physics and assist to obtain intriguing strongly correlated states hardly accessible otherwise. We demonstrated that the feedback control and optical measurement backaction can be used for engineering quantum states with long-range correlations which can be tuned by changing the spatial structure of the jump operator. We illustrated this by considering the case where the measurement induces two macroscopically occupied spatial modes and the feedback loop can stabilize interesting quantum phases such as supersolid-like states (i.e. the density waves with long-range matter-wave coherence) and states with antiferromagnetic correlations, depending on the value of the feedback gain $z$. This parameter determines the strength of the feedback loop and its net effect is to modulate the tunneling amplitude $J$ according to the measurment outcome. Experimentally, this can be achieved by changing the depth of the optical lattice $V_0$  in \cite{Buchler2003}
\begin{equation}
J=\frac{4 E_R}{\sqrt{\pi}}\left( \frac{V_0}{E_R}\right)^{3/4}\exp \left( - 2 \sqrt{\frac{V_0}{E_R}}\right)
\end{equation}
where  $E_R$ the recoil energy. Importantly, since the critical value of $z$ scales as $1/(\gamma N^2)$ the effects described in this work can be observed by weakly perturbing  $V_0$, changing the optical lattice depth by slightly modifying the intensity of the lasers. For example, assuming $V_0/E_R\sim 10$ and  considering $z\sim 10^{-2}$, one needs to change the ratio $V_0/E_R$ by a factor $\sim 5 \cdot 10^{-3}$.  Moreover, the light-atom coupling regime we considered has been recently realized in two different experimental setups \cite{Hemmerich2015,Esslinger2015}. Finally, these phenomena are not specific to ultracold atomic systems but the same ideas can be applied to superconducting qubits \cite{Palacios2010,Paraoanu2011,Pirkkalainen2013,White2015}, molecules \cite{MekhovLP2013}, optomechanical arrays \cite{Paternostro2011,Aspelmeyer2014}, arrays of optical resonators \cite{Minkov:16}, Rydberg \cite{Zeiher2016,Schauss2015} and other polaritonic and spin excitations \cite{Budroni2015,Angelatos:16} 
and even purely photonic systems with multiple path interference \cite{Spring2013,Nitsche2016,Brecht2015,Elster2015,Oren:16}. Moreover, they can be extended for the cases of light-matter interaction with conventional condensed matter systems with strong correlations (e.g. high-Tc superconductors) \cite{Mitrano,Zhang2016}.

\section*{Funding Information}
The work was supported by the EPSRC (EP/I004394/1).

%

\end{document}